\documentclass{elsart}
\usepackage{natbib}
\usepackage{graphicx}
\begin{document}
\runauthor{Olivier Ravel}
\begin{frontmatter}
\title{Radio Detection of Cosmic Ray Air Shower by the CODALEMA Experiment}

\author[Nantes]{O.Ravel}\footnote{Email : Olivier.Ravel@subatech.in2p3.fr}, 
\author[Nantes]{R.Dallier}, 
\author[Nancay]{L.Denis},
\author[Nantes]{T.Gousset},
\author[Nantes]{F.Haddad},
\author[Nantes]{P.Lautridou},
\author[Meudon]{A.Lecacheux},
\author[Nantes]{E.Morteau}, 
\author[Meudon]{C.Rosolen},
\author[Nantes]{C.Roy} 

\address[Nantes]{SUBATECH, Universit\'e de Nantes/In2p3-CNRS/Ecole des Mines de
Nantes \\ 4 rue A.Kastler, 44307 Nantes cedex 3, France}
\address[Nancay]{Station de Radioastronomie de Nan\c{c}ay \\ 18330 Nan\c{c}ay, France}
\address[Meudon]{LESIA, Observatoire de Paris-Meudon \\ place Jules Janssen, 92195 Meudon, France.}

\begin{abstract}
The  possibilities of measuring Extremely High Energy Cosmic Rays (EHECR)  
by radio detection of  electromagnetic  pulses radiated during the development
of extensive air showers in the atmosphere are investigated. 
We present the demonstrative CODALEMA experiment, set up at Nan\c{c}ay Radio-Observatory (France). 
The radio-decametric array has been adapted to measure radio transients in time coincidence between antennas.
\end{abstract}
\end{frontmatter}

\section {Motivation}
\indent Radio emission from cosmic ray air showers has been predicted in the 60's~\cite{askaryan}. In the subsequent decade, several experiments have
asserted the observation of such emission~\cite{jelley,Weeks}.
For this period, following the review of Allan~\cite{Allan}, different mechanisms of electromagnetic pulse generation have been identified,
such as negative charge excess (10-25\%) in the shower.
For our purpose, calculations of the electromagnetic pulse were performed within a simple model considering only this negative charge excess contribution. 
The resulting electric field  can be expressed as :
$$\vec{E}(t)=\frac{1}{4\pi\epsilon}\sum_{t'} \frac{e(t')(1-\beta^2)(\vec{n}-\vec{\beta})}{R^2 |1-\vec{n}\cdot\vec{\beta}|^3} +
\frac{1}{4\pi\epsilon c}\sum_{t'} \frac{e'(t')(\vec{n}-\vec{\beta})}{R(1-\vec{n}\cdot\vec{\beta})|1-\vec{n}\cdot\vec{\beta}|}  $$

with $\vec{\beta}=\vec{v}/c$, $\vec{n}=\vec{R}/R$, and $c(t-t')=R$ the distance between the moving charges of the shower and the
observation point. For the numerical simulation, we have assumed a charge excess distribution $e(t')$ along the shower axis with
 $7.10^9$ charges at the maximum for a 100 EeV  EHECR and no spatial shower extension.
The corresponding electric fields for 3 different  impact parameters $b$ in a vertical shower configuration are shown on Fig.1.
\begin{figure}
\begin{center}
\includegraphics[width=8.cm]{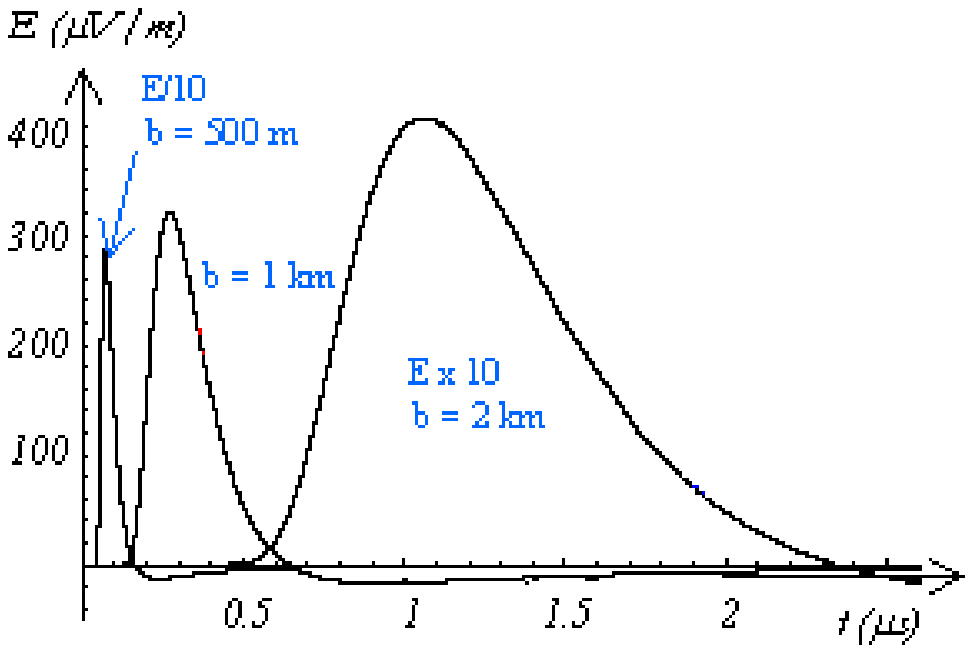}
\caption{ \it Simulated electric field as a function of time for impact parameter $b=0.5,~1~and~2~km$.} 
\end{center}
\begin{center}
\includegraphics[width=8.cm]{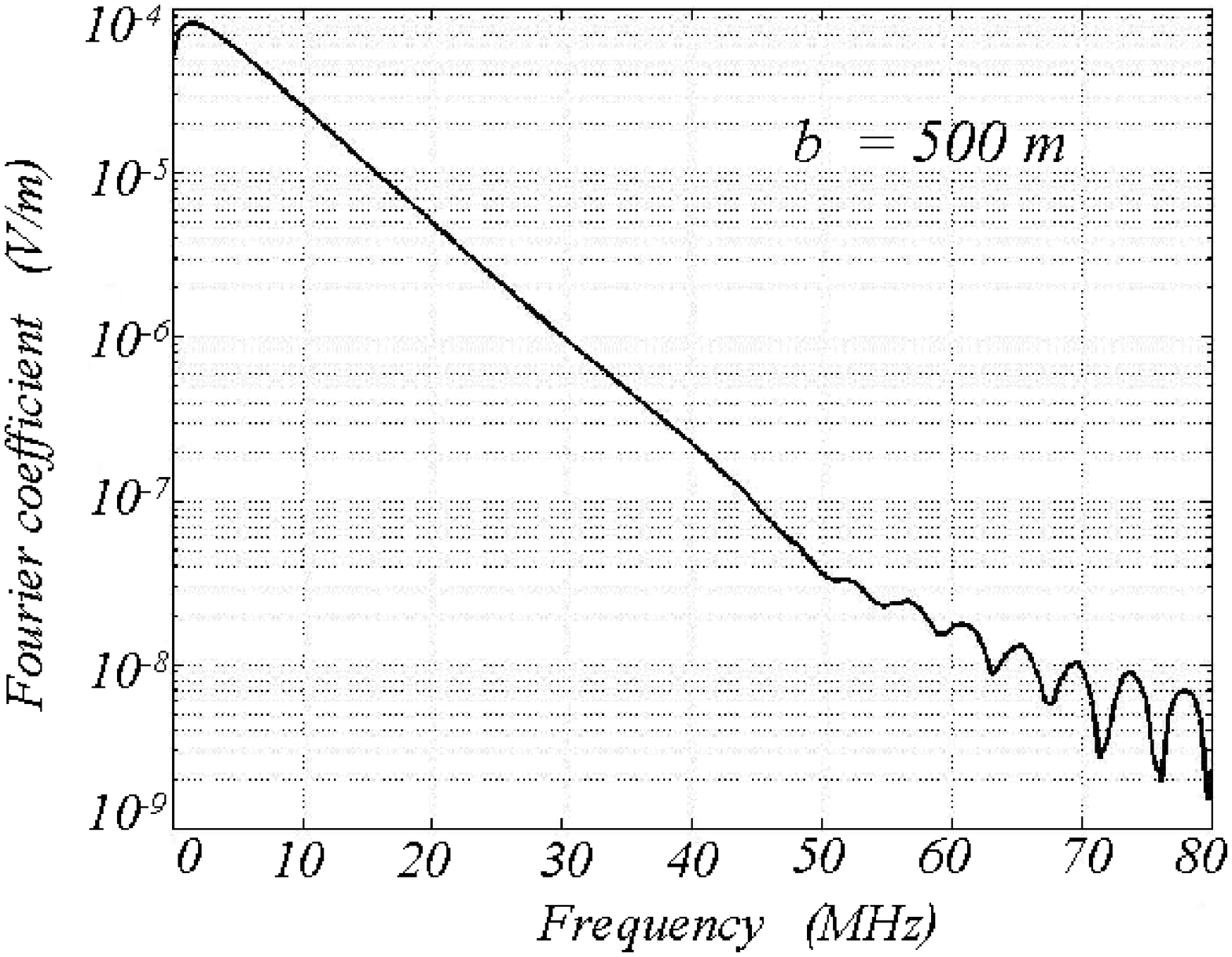}
\caption{\it Fourier spectrum of a simulated pulse at $b= 500~m$} 
\end{center}
\end{figure} 
These electric fields are strong enough to be measured. It is worth noting that the pulse width increases with the impact parameter. 
Thus, a complete knowledge of the pulse  waveform enables to determine the charge profile of the shower and 
consequently the energy  of the primary EHECR.
Due to the large frequency content of the  Fourier spectrum (see Fig. 2), broadband  antennas are requested for the most accurate pulse shape
determination. It is also possible, by  an appropriate narrow-band frequency filter, to use stand-alone triggered antennas. 
Moreover, using at least 3 antennas in time coincidence enable to reconstruct the shower direction.
\section{The CODALEMA experiment}

\begin{figure}
\begin{center}
\includegraphics[width=\linewidth]{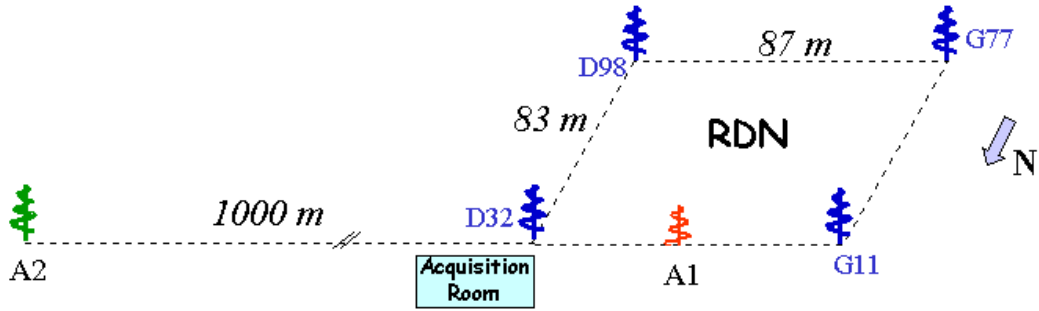}
\caption{\it CODALEMA setup} 
\end{center}
\end{figure} 

These considerations motivated us to implement a demonstrative experiment, named CODALEMA, at the Radio Observatory of Nan\c{c}ay~\cite{nancay}
(France). CODALEMA uses 6 large frequency bandwidth antennas (1-100 MHz) in time coincidence.
The experimental central array (83 m x 87 m) consists of the corner antennas of the R\'eseau D\'ecam\'etrique de Nan\c{c}ay (RDN) 
and a narrow-filtered antenna (33-65 MHz) used as a trigger. One additional antenna linked by optical fiber is located 1 km East from the
central array (see Fig. 3).

The data acquisition system is based on 3 LeCroy digital oscilloscopes (1 GS/s) which record the amplified antenna voltage signal.
Our first task has been to identify the various sources of background, mainly the radio broadcasting stations. Bandpass filters (24-82 MHz)
suppress the AM and FM bands. Additional anthropic noise comes from  the acquisition room's air-conditioning
and the computing network hub : these effects have been filtered out.
Within these precautions we have reached a trigger rate of 12 events/hour for a 40 $\mu$V triggering antenna signal; among them 85\% have
 been already identified as noise pulses.
 
\section {First results}
A typical 6 antennas coincidence signal, triggered by the dedicated narrow filtered antenna of the central array is shown Fig. 4. After a preliminary
analysis of time delays, we can assert that such an event is not compatible with identified sources of noise and can be considered as a possible
candidate for cosmic ray air shower radio emission.
CODALEMA has been now recording data at a duty cycle of 100\% since mid-march 2003, and until the end of 2003. 

\begin{figure}
\begin{center}
\includegraphics[width=\linewidth]{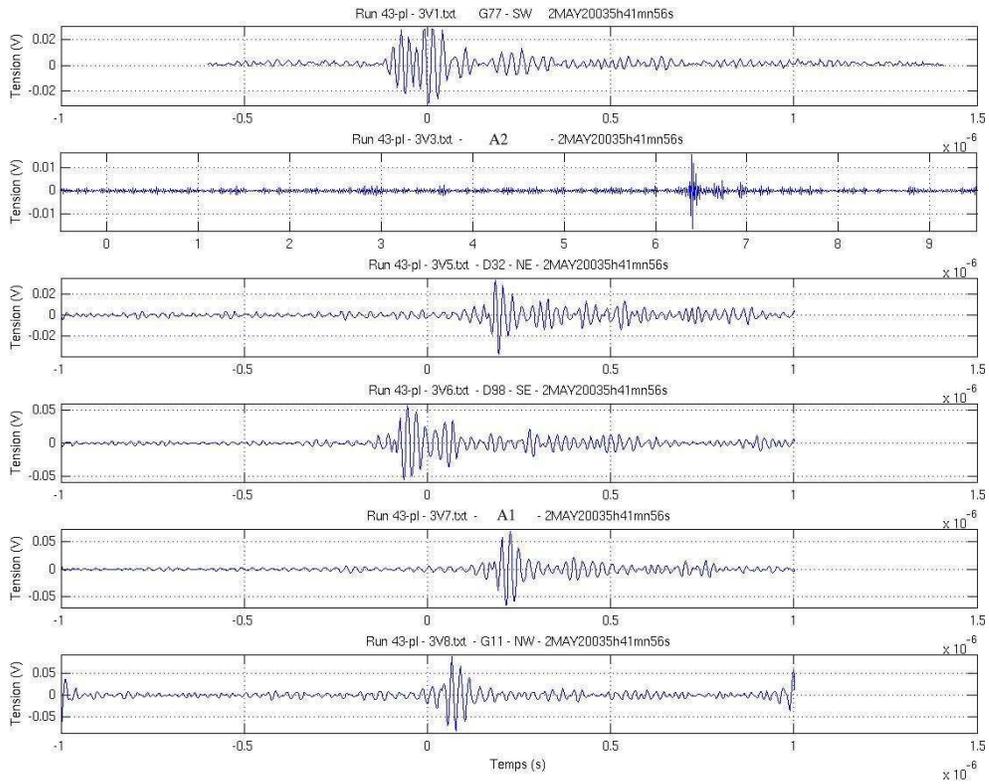}
\caption{ \it Coincidence transients after numerical filtering (35-65 MHz) detected with CODALEMA} 
\end{center}
\end{figure}

{\em Acknowledgments :} We thank J.Martino and N.Dubouloz for their full supports.

\end{document}